# Comparison of diffuse correlation spectroscopy analytical models for cerebral blood flow measurements

**Mingliang Pan*, Quan Wang, Yuanzhe Zhang, and David Day-Uei Li***

University of Strathclyde, Department of Biomedical Engineering, Glasgow G4 0NW, United Kingdom

**Abstract**

**Significance:** Although multi-layer diffuse correlation spectroscopy (DCS) analytical models have been proposed to reduce contamination from superficial signals when probing cerebral blood flow index (CBFi), a comprehensive comparison and clear guidance for model selection remains lacking. This report aims to address this gap.

**Aim:** This study aims to systematically compare three DCS analytical models: the semi-infinite, two-layer, and three-layer models, with a focus on their fundamental differences, data processing approaches, and the accuracy and reliability of CBFi estimation. We also provide practical recommendations for selecting the most appropriate model based on specific application scenarios, to support researchers in applying DCS effectively.

**Approach:** Experimental data were generated by simulating a four-layer slab head model using the Monte Carlo eXtreme (MCX) toolkit. We evaluated various fitting strategies for three DCS models; early time lag range (ETLR) fitting with or without treating the coherence factor $\beta$ as a fitting parameter for the semi-infinite model; single-distance (SD) and multi-distance (MD) fitting for the two- and three-layer models. We then compared their performance in terms of CBF sensitivity, recovery of relative CBFi (rCBFi) changes, accuracy of absolute CBFi estimates across different source-to-detector separations ($\rho$ = 20, 25, 30, 35 mm), ability to separate the crosstalk from extracerebral layers (scalp BFi (SBFi), and skull BFi (BBFi)), sensitivity to parameter assumption errors, and time-to-result, using the respective optimal fitting strategies for each model.

**Results:** The optimal fitting methods for estimating CBFi are: ETLR fitting with a constant $\beta$ for the semi-infinite model; SD fitting with $\beta$ fixed for the two-layer model; and MD fitting for the three-layer model. The two-layer and three-layer models exhibit enhanced CBFi sensitivity, approaching 100%, compared to 36.8% for the semi-infinite model at $\rho$ = 30 mm. The semi-infinite model is suitable only for rCBFi recovery at a larger $\rho$ ($\geq$ 30 mm). In contrast, the two-layer model is appropriate for both CBFi and rCBFi recovery across all tested $\rho$ values (20, 25, 30, 35 mm in this work), although its robustness declines as $\rho$ increases. The three-layer model enables simultaneous recovering of CBFi, SBFi, and rCBFi. Among these, the two-layer model is the most effective at mitigating the influence of extracerebral BFi, whereas CBFi estimates from the semi-infinite and three-layer models remain consistently affected by variations in SBFi and BBFi. Errors in assumed model parameters have minimal impact on rCBFi recovery across all models. In terms of computational efficiency, the semi-infinite model requires only 0.38 seconds of processing 500 data samples, demonstrating potential for real-time rCBFi inference. In comparison, the two-layer and three-layer models require substantially longer processing times of 9,502.18 seconds and 35,099.34 seconds, respectively.

**Conclusions:** This systematic comparison of three DCS analytical models demonstrates the superior ability of multi-layer models to reduce the influence of superficial tissue layers, thereby enhancing CBFi and rCBFi sensitivity relative to the semi-infinite model. We evaluated various fitting strategies and, beyond recommending the optimal approach for each model, we provide practical guidance for selecting the most appropriate model based on specific objectives, experimental conditions, and data analysis requirements. We believe this work offers a valuable reference for researchers in the field, supporting informed model selection and highlighting key considerations for the effective application of DCS analytical models.

*Mingliang Pan, E-mail: mingliang.pan@strath.ac.uk;  David Day-Uei Li, E-mail: david.li@strath.ac.uk

# 1 Introduction

Blood flow (BF) is a critical clinical biomarker of human health, responsible for transporting oxygen, nutrients, and metabolic byproducts through the bloodstream[1]. Diffuse correlation spectroscopy (DCS) is a non-invasive optical technique that uses near-infrared light to assess deep tissue BF[2]. It quantifies BF by calculating the normalized intensity autocorrelation function (ACF, $g_2$) of fluctuations in scattered coherent light , primarily arising from the motion red blood cells (RBCs)[3], and offers a high temporal resolution ranging from 1 to 100 Hz[4–6]. DCS has been broadly applied to monitor brain health/functions[7–9], cardio-cerebrovascular diseases[10–13], neurovascular coupling[14,15], and tumor diagnosis[16] and therapy[17].

Traditionally, the measured $g_2$ is fitted to an analytical solution assuming a semi-infinite homogeneous tissue model to extract the BF index (BFi)[18,19]. However, this simplified model neglects the anatomical complex of human tissue. For example, it is insufficient for analyzing cerebral BFi (CBFi), as the human head comprises multiple layers, including scalp, skull, cerebrospinal fluid, grey matter, and white matter[20,21]. Due to the tissue heterogeneity, DCS signals originating from deeper brain layers can be contaminated by the BF in superficial layers (i.e. scalp and skull)[22–28], often leading to underestimation of CBFi[24,25,29].

Various approaches have been proposed to enhance the accuracy of CBFi measurement, including early time lag range (ETLR) fitting, pressure modulation[30], application of correction factors[25], depth-sensitive time-domain DCS (TD-DCS)[31–33], interferometric DCS (iDCS)[34,35], long wavelength DCS[36], and multi-channel detection[37,38]. In parallel, analytical models have evolved from the traditional semi-infinite model to more advanced multi-layer models, including two-layer and three-layer models[4,25,39–41]. Comparative studies have shown that multi-layer models offer advantages over the semi-infinite model in estimating relative changes in CBFi, whereas the semi-infinite model remains more sensitive to signals from superficial layers[4,22,23,25,28,42].

However, comprehensive comparisons between different analytical models remain lacking. Although the two-layer and three-layer models have been compared with the semi-infinite model[23,28,42,43], direct comparisons involving all three models are limited. Zhao et al.[44] conducted a direct comparison focusing on CBFi extraction using a clinical collected dataset; however, they did not address differences in relative CBFi (rCBFi) recovery, the influence from extracerebral layers, or the impact of errors in assumed model parameter. Table 1 summarizes the contributions of various studies comparing analytical models and highlights the unique contributions of this work.

Although previous studies have explored different models, they did not provide a quantitative comparison of crosstalk from extracerebral layers, nor did they examine crosstalk specifically originating from the skull layer BFi (BBFi). Overall, existing literature compares models in a fragmented manner, lacking comprehensiveness and offering no guidance on model selection for different experimental scenarios. Furthermore, optimal fitting strategies for these models remain unclear, and comparisons between existing fitting methods, such as single-distance (SD) and multi-distance (MD) fitting, are still needed. This article aims to address these gaps by presenting a systematic comparison of the three analytical models.

**Table 1** Summary of previous studies that compared DCS analytical models.

| Studies | Models | Data source | $\rho$ (mm) | CBF sensitivity | Crosstalk from extracerebral BFi | Optical and physiological parameters assuming errors | Fitting methods |
|---|---|---|---|---|---|---|---|
| **Gagnon et al.**[4] (2008) | Semi-infinite Two-layer | Two-layer slab MC simulations; MRI-scanned head model simulations; Two-layer liquid phantom | 10, 20, 30 | ✗ | ✗ | ✓ | Whole curve SD fitting |
| **Verdecchia et al.**[25] (2016) | Semi-infinite Three-layer | Two-layer liquid phantom; Pig head experiment | 20, 27 | ✗ | ✗ | ✗ $\mu_a, \mu_s'$: TR-NIRS Thickness: CT | Semi-infinite: ETLR fitting; Three-layer: MD fitting |
| **Wu et al.**[22] (2021) | Semi-infinite MC three-layer model | MC simulations; Human subject with hypercapnia | 5, 25, 30 | ✗ | ✓ | $\mu_a, \mu_s'$: Assumed Thickness: MRI | Semi-infinite: ETLR fitting |
| **Zhao et al.**[42] (2021) | Semi-infinite Three-layer | Three-layer analytical model; MC human head simulations | 10, 25 | ✗ | ✗ | ✓ | Semi-infinite: / Three-layer: MD fitting |
| **Forti et al.**[23] (2023) | Semi-infinite Two-layer | Three-layer slab MC simulations; Two-layer liquid phantom | 25 | ✗ | ✗ | ✓ | Semi-infinite: ETLR fitting |
| **Zhao et al.**[44] (2023) | Semi-infinite Two-layer Three-layer | Clinical dataset from patients with subarachnoid hemorrhage | 10, 25 | ✓ | ✗ | ✗ $\mu_a, \mu_s'$: Assumed Thickness: CT | Semi-infinite: ETLR fitting; Two-layer: MD fitting; Three-layer: MD fitting |
| **Wang et al.**[28] (2023) | Semi-infinite Three-layer | Three-layer slab MC simulations | 5, 10, 15, 20, 25, 30 | ✓ | ✗ | ✓ | Whole curve SD fitting |
| **This work** | Semi-infinite Two-layer Three-layer | Four-layer slab MC simulations | 20, 25, 30, 35 | ✓ | ✓ | ✓ (short review) | Semi-infinite: ETLR fitting; Two-layer: SD fitting; Three-layer: MD fitting |

In this work, we simulate a four-layer slab human head model to generate experimental data. We investigate different fitting strategies for the three analytical models and explore their intrinsic CBFi sensitivity. We compare the accuracy of each model in recovering absolute CBFi and rCBFi across various source-detector separations ($\rho$ = 20, 25, 30, and 35 mm). We also assess how CBFi and rCBFi accuracy is influenced by assumptions regarding optical properties and the thickness of the scalp and skull, as well as by each model's robustness to blood flow changes in extracerebral layers. In addition, we evaluate the time-to-result for each model using its respective fitting method, to assess the feasibility of real-time CBFi monitoring. Finally, we discuss several key findings and current limitations of the study, and provide recommendations for selecting the most appropriate model for specific applications and experimental conditions.

## 2 Methods

*2.1 DCS theory*

The DCS theory is based on the correlation diffusion equation (CDE), derived from correlation transfer equation (CTE) under the standard diffusion assumption[45,46]. This derivation is analogous to the photon diffusion equation (PDE) from the radiative transfer equation (RTE) using the $P_N$ approximation[46,47]. The analogy between the CTE to RTE was firstly established by Ackerson *et al.*[48]. The CDE is expressed as:

$$\left(\frac{D_r}{v}\nabla^2 - \mu_a - \frac{1}{3}\mu_s' k_0^2 \alpha \langle \Delta r^2(\tau)\rangle\right) G_1(\mathbf{r},\tau) = -S(\mathbf{r}), \quad (1)$$

where $G_1(\mathbf{r},\tau) = \langle E(r,t)E^*(r,t+\tau)\rangle$ is the electric field temporal ACF, $D_r = v/(3\mu_s')$ is the photon diffusion coefficient, $v$ is the speed of light in the medium, $\tau$ is the delay time, $k_0 = 2\pi n_0/\lambda$ is the wavenumber of light in the scattering medium at the wavelength $\lambda$ and $n_0$ is the tissue refractive index, $\mu_a$ is the absorption coefficient, $\mu_s'$ is the reduced scattering coefficient, $S$ is the source, and $\langle \Delta r^2(\tau)\rangle$ is the mean square displacement of scatterers. For diffusive motions, $\langle \Delta r^2(\tau)\rangle = 6D_B\tau$, where $D_B$ is the effective Brownian diffusion coefficient. In most practical applications, the Brownian motion model is accurate to describe scatterers' motions[19,49,50]. The product $\alpha D_B$ is defined as BFi[51], where $\alpha$ is the probability of scattering from a moving scatterer, assumed to be 1 in our simulations[52].

The normalized electric field temporal ACF, $g_1(\tau)$, is related to the normalized light intensity ACF, $g_2(\tau)$, through the Siegert equation[53]:

$$g_2(\tau) = 1 + \beta|g_1(\tau)|^2, \quad (2)$$

where $\beta$ is the coherence factor, depends on the laser stability, coherence length, and the number of speckles detected[19]. The measured light intensity ACF can be calculated as:

$$g_2(\tau) = \frac{\langle I(t)I(t+\tau)\rangle}{\langle I(t)\rangle^2}, \quad (3)$$

where $\langle \dots \rangle$ denotes the average over the integration time $T_{int}$, and $I(t)$ is the measured light intensity fluctuation. By fitting Eq. (3) to the analytical light intensity ACF, the BFi ($\alpha D_B$) can be extracted.

In this study, we consider light propagation using a commonly used continuous-wave (CW) 785 nm laser source isotropically scattered in the medium. The Green's function solution to Eq. (1), $G_1(\mathbf{r},\tau)$, has different forms under semi-infinite, two-layer, and three-layer boundary conditions. Detailed derivations of DCS analytical models can be found in Supplementary Material.

*2.2 Data source*

In this work, experimental data were generated using Monte Carlo (MC) simulations conducted with the voxel-based Monte Carlo eXtreme (MCX)[54] toolkit in MATLAB (R2023b, The MathWorks). We simulated a slab human head model, with a volume of 200×200×200 mm³, segmented into four layers, each layer represent scalp, skull, cerebrospinal fluid (CSF), and brain tissues. 785 nm light was used in all MC simulation in this work, and the optical parameters at 785 nm are detailed in Table 2. Tissue refraction indices were assumed as 1.37 for all layers. Each simulation was executed with $10^9$ photons from a 1 mm diameter source and utilized five detectors (1 mm in diameter, positioned at $\rho$ = 15, 20, 25, 30, 35 mm) to simultaneously record the photon transfer and photon pathlength, thereby enabling the calculation of the temporal light field ACF[51,55]:

$$G_1(\tau) = \frac{1}{N_p}\sum_{n=1}^{N_p} \exp\left(\sum_{i=1}^{N_{tissue}} -\frac{1}{3}Y_{n,i}k_0^2 \langle \Delta r^2(\tau)\rangle_i\right) \exp\left(-\sum_{i=1}^{N_{tissue}} \mu_{a,i}L_{n,i}\right), \quad (4)$$

where $N_p$ is the number of detected photons at each detector, $N_{tissue}$ is the number of tissue types ($N_{tissue}$ = 4 in our case), $Y_{n,i}$ and $L_{n,i}$ are the total momentum transfer and total pathlength of photon $n$ in Layer $i$, and $\mu_{a,i}$ is the absorption coefficient in Layer $i$. $\langle \Delta r^2(\tau)\rangle_i = 6D_{Bi}\tau$ (see Sec. 2.1) is the mean square displacement of the scattered particles in Layer $i$, where $D_{Bi}$ is the effective Brownian diffusion coefficient in Layer $i$. The correlation delay time $\tau$ was adopted from our time-tagger module (SPC-QC-104, Beker & Hickl). The simulated $G_1(\tau)$ was normalized to $G_1(0)$, then $g_2(\tau)$ curves were obtained through Eq. (2) with $\beta = 0.5$.

Table 2 Optical parameters at 785 nm and layer thicknesses of the four-layer human head model simulation.

| | Layer | Layer thickness (mm) | $\mu_a$ (mm$^{-1}$) | $\mu_s$ (mm$^{-1}$) | g | $D_B$ (mm$^2$/s) |
|---|---|---|---|---|---|---|
| **four-layer slab** | Scalp | 5 | 0.019 | 6.600 | 0.89 | $1 \times 10^{-6}$ |
| | Skull | 7 | 0.014 | 8.600 | | $8 \times 10^{-8}$ |
| | CSF | 2 | 0.001 | 0.002 | | $1 \times 10^{-8}$ |
| | Brain | ∞ | 0.020 | 11.00 | | $6 \times 10^{-6}$ |

We use the slab head model because it allows convenient comparison of the CBFi sensitivity by adjusting $\rho$, and previous research has validated that the slab model provides a reasonable approximation of the anatomically curved human head[22]. It is known that the diffuse approximation becomes invalid in regions of low scattering, such as the CSF[21,54,56]. Despite this, studies by Custo et al.[57] and Zhao et al.[6] indicate that the CSF has only a minor influence on brain sensitivity and rCBFi recovery. To better represent a realistic human head in our simulation, we included a 2mm-thick CSF layer as reported by Wu et al.[58] and Okada et al[59].

The simulation parameters, summarized in Table 2, are based on biologically realistic tissue properties reported in previous studies at a wavelength of 785 nm[4,21,60–63]. The reduced scattering coefficient $\mu_s'$ is calculated using $\mu_s' = (1-g)\mu_s$, where $\mu_s$ is the scattering coefficient and $g$ is the tissue's scattering anisotropy factor[64]. To more accurately mimic the human head's layered structure, we set the scalp and skull thicknesses to 5 mm and 7 mm, respectively, based on prior studies[57,65,66].

The baseline Brownian diffusion coefficients $D_{Bi}$ ($i$ = 1, 2, 3, 4) were set as 1×10$^{-6}$, 8×10$^{-8}$, 1×10$^{-8}$, 6×10$^{-6}$ mm$^2$/s for scalp[36,67], skull, CSF[6], and brain layers[28,58], respectively. Notably, direct measurements of skull blood flow are limited [30,36]. However, the skull is known to contain vascular networks, whereas the CSF layer primarily contains small perivascular capillaries[68]. Therefore, we simulate minimal blood flow in the skull and negligible flow in the CSF layer. The simulated $g_2$ data from the four-layer model served as the experimental data for our analysis.

*2.3 Noise model*

To mimic real-world DCS measurements and assess the robustness of the analytical models, we employed a Gaussian noise model with zero mean and a standard deviation based on the noise model proposed by Zhou *et al.*[69] to our simulations. The standard deviation is defined as:

$$\sigma(\tau) = \sqrt{\frac{T_{\text{bin}}}{T_{\text{int}}} \left[ \beta^2 \frac{(1 + e^{-2\Gamma T_{\text{bin}}})(1 + e^{-2\Gamma \tau}) + 2m(1 - e^{-2\Gamma T_{\text{bin}}})e^{-2\Gamma \tau}}{1 - e^{-2\Gamma T_{\text{bin}}}} + \frac{2\beta(1 + e^{-2\Gamma T_{\text{bin}}})}{\langle n \rangle} + \frac{1 + \beta e^{-\Gamma \tau}}{\langle n \rangle^2} \right]^{1/2}}, \quad (5)$$

where $\Gamma$ is the decay speed, extracted by fitting the simulated $g_2(\tau)$ curves with a single-exponential decay model: $g_2(\tau) = 1 + \beta \exp(-2\Gamma \tau)$[69]. $T_{\text{bin}}$ is the correlator bin width, adopted from our time-tagger module. It is set to be 6.145 ns for the first 16-channel and is tripled every 16 channels thereafter. $T_{\text{int}}$ is the integration time, $m$ is the bin index, $\langle n \rangle$ is the average number of photons within the bin width $T_{\text{bin}}$, where $\langle n \rangle = I T_{\text{bin}}$ And $I$ is the detected photon count rate. Gaussian noise with zero mean and a standard deviation defined by Eq. (5) was added to clean $g_2$ data to simulate realistic testing conditions.

The photon count rate and the integration time can influence the noise level in DCS measurements. Generally, a higher photon count rate and longer integration time yield a higher signal-to-noise ratio (SNR). However, in practice, the photon count rate is constrained by safety regulations governing maximum permissible light exposure on biological tissues, as specified in ANSI Z136.1[70], and it is typically not a controllable parameter for a specific $\rho$. In contrast, the integration time can be adjusted according to the sampling rate[5].

In this work, we assumed a photon count rate $I$ of 10 kcps at $\rho = 30$ mm, consistent with previous reports[36]. Photon count rates at other $\rho$ were estimated by comparing the Green's function solution for photon fluence at delay time zero, $G_1(0)$[71]. The integration time $T_{\text{int}}$ was set to 60 s. The SNR of the synthesized $g_2$ was calculated using[72,73]:

$$\text{SNR} = (g_2(\tau) - 1)/\sigma(\tau), \quad (6)$$

We have validated the noise model using measurements on a milk phantom; the validation results are provided in the Supplementary Material, Fig. S2. For our simulations, 20 noise realizations were added to each CBFi condition. The resulting synthetic noisy dataset was used to evaluate the performance of the analytical models under realistic experimental conditions.

*2.4 Fitting methods*

For the semi-infinite analytical model, previous studies have shown that early time lag range (ETLR) fitting can enhance CBF sensitivity, although this improvement comes at the cost of reduced SNR[24,58,74]. For the three-layer model, Zhao *et al.*[42] and Verdecchia *et al.*[25] employed MD fitting to obtain optimized solutions. In this study, we apply ETLR fitting to the semi-infinite model and evaluate both MD and single-distance (SD) fitting approaches for the multi-layer models.

<u>Penalty function for the semi-infinite fitting:</u>

For ETLR fitting, previous studies have constrained $g_2 \geq 1.25$[22,75]. In this work, to facilitate our analysis, we simply restrict the ETLR to $\tau \leq 30$ $\mu s$ across different $\rho$[76] as shown in Fig. 1. We use *fminsearchbnd* function in MATLAB to minimize the single penalty function defined as:

$$\chi^2 = \sum_{i=1}^{N_\tau} \left[ g_2^{\text{theory}}(\rho, \tau_i, \text{CBFi}_{\text{estimated}}) - g_2^{\text{simulated}}(\rho, \tau_i, \text{CBFi}_{\text{true}}) \right]^2, \quad (7)$$

where $N_\tau$ is the number of $\tau$, and $\tau_i$ is the $i$'th delay time. CBFi is the fitting parameter, constrained within the bounds $\in [10^{-9}, 10^{-4}]$ mm$^2$/s. We extracted CBFi at $\rho$ = 15, 20, 25, 30, 35 mm. The optical parameters for the semi-infinite fitting were assumed to be known and taken from the brain layer, as listed in Table 2. In previous studies, CBFi and $\beta$ were simultaneously estimated through fitting[28,77]. However, treating $\beta$ as a fitting parameter tends to reduce the convergence and stability of the estimated CBFi. In this work, we compare semi-infinite ETLR fitting with and without setting $\beta$ as a fitting parameter.

As the use of single-exponential fitting for processing DCS data to estimate rCBFi has become increasingly common, it is important to note that the reliability of this method has not yet been validated against the semi-infinite analytical fitting. It is known that the Green's function solution to the CDE under a semi-infinite geometry can be simplified to a single-exponential decay function at short correlation times[69,78,79], as is introduced in Sec. 2.3. Under this approximation, the normalized electric field ACF is expressed as: $g_1(\tau) = e^{-\tau/\tau_c}$. According to Siegert's equation[53], the normalized light intensity ACF is then given by $g_2(\tau) = 1 + \beta |g_1(\tau)|^2$, i.e.,

$$g_2(\tau) = 1 + \beta e^{-\frac{2\tau}{\tau_c}}, \quad (8)$$

where $\tau_c$ is the decorrelation time. In this work, we also compare the single-exponential fitting using Eq. (8) with semi-infinite ETLR analytical fitting. The penalty function for single-exponential fitting is defined as:

$$\chi^2 = \sum_{i=1}^{N_\tau} \left[ g_2^{\text{exponential}}(\rho, \tau_i, \tau_c) - g_2^{\text{simulated}}(\rho, \tau_i, \text{CBFi}_{\text{true}}) \right]^2, \quad (9)$$

The reciprocal of $\tau_c$, referred to as the decorrelation speed, is used to calculate rCBFi (rCBFi = $\tau_{c\_\text{baseline}}/\tau_c$). *fminsearchbnd* function was used to to minimize the penalty function in MATLAB. This result was then compared with rCBFi recovered from the semi-infinite ETLR analytical fitting.

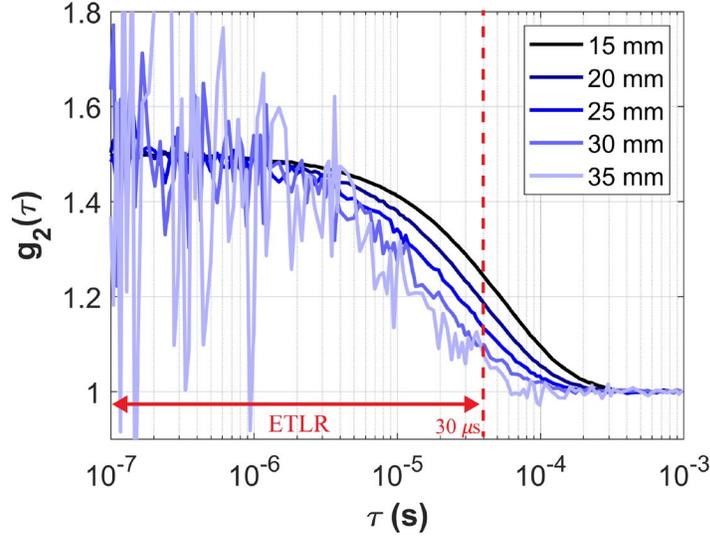

**Fig. 1** $g_2$ curves at different $\rho$, the red line with arrows shows the ETLR window.

Penalty function for multi-layer models fitting:

In addition to fitting a single $g_2$ curve, we also implemented MD fitting for the two- and three-layer models. Namely, we simultaneously fit two $g_2$ curves, one is at a smaller $\rho$ ($\rho$ = 15 mm), and another is a larger $\rho$ ($\rho$ = 20, 25, 30, 35 mm). MD fitting has been demonstrated to improve CBFi recovery accuracy when using multi-layer DCS models[55]. The penalty function for multi-layer models fitting is defined as:

$$\chi^2 = \sum_{j=1}^{N_\rho} \sum_{i=1}^{N_\tau} \left[ g_2^{\text{theory}}(\rho_j, \tau_i, \text{SBFi}_{\text{recovered}}, \text{CBFi}_{\text{recovered}}) - g_2^{\text{simulated}}(\rho_j, \tau_i, \text{SBFi}_{\text{true}}, \text{CBFi}_{\text{true}}) \right]^2, (10)$$

where $\rho_j$ is the $j$'th $\rho$ and $N_\rho$ = 1 for SD multi-layer fitting (= 2 for MD multi-layer fitting). SBFi and CBFi are the fitting parameters, we use *fminsearchbnd* function to minimize $\chi^2$, and set the bounds for SBFi and CBFi both $\in [10^{-9}, 10^{-4}]$ mm$^2$/s. As in previous cases, the optical and physiological parameters (e.g. layer thicknesses) were assumed to be known prior to fitting, as detailed in Table 2. For the three-layer fitting, the first and second layers represent the scalp and skull layers of the four-layer slab head model. The CSF layer is neglected and grouped with the cerebral layer, following the approach reported in a previous study[6]. For the two-layer fitting, the extracerebral optical parameters are taken from the scalp layer of the four-layer model, and the extracerebral layer thickness is defined as the combined thickness of the scalp and skull layers, again, with the CSF included in the cerebral layer.

For MD fitting $\beta$ was treated as a fitting parameter with bounds $\in [0, 1]$; for SD fitting, $\beta$ was either fixed at 0.5 or treated as a fitting parameter. We compared MD and SD fitting strategies for the multi-layer analytical models, and based on this comparison, selected the optimal fitting approach for CBFi recovery in subsequent assessments using the multi-layer models.

*2.5 CBFi sensitivity, and absolute CBFi, rCBFi recovery*

To evaluate the sensitivity of different analytical models to changes in CBFi, we simulated CBF perturbations in Layer 4 of the four-layer model. Specifically, $D_B$ in Layer 4 (brain layer) was varied by ±25% and ±50% relative to the baseline ($D_{B0} = 6 \times 10^{-6}$ mm²/s), while keeping $D_B$ in Layers 1, 2, 3 constant (i.e., $D_{B1,2,3} = 1 \times 10^{-6}, 8 \times 10^{-8}, 1 \times 10^{-8}$ mm²/s). As described in Sec. 2.3, 20 noise realizations were added to each CBFi condition, resulting in 100 samples for each $\rho$ value (20 realizations × 5 conditions, including one baseline and four perturbation levels). The mean CBFi value recovered under the baseline condition is used as the global baseline for calculating rCBFi, defined as[80]:

$$\text{rCBFi} = \frac{\text{CBFi}_{\text{recovered}}}{\text{CBFi}_{0\_\text{recovered}}}, \tag{11}$$

where $\text{CBFi}_{\text{recovered}}$ and $\text{CBFi}_{0\_\text{recovered}}$ are the recovered CBFi at perturbed and baseline conditions, respectively. The generated perturbed $g_2$ data was then used to extract absolute SBFi and CBFi, rCBFi and calculate the CBFi sensitivity defined as[24,62]:

$$S_{\text{CBFi}} = \frac{\left(\text{CBFi}_{\text{recovered}}/\text{CBFi}_{0\_\text{recovered}}\right) - 1}{\left(D_{B4}/D_{B0,4}\right) - 1} \times 100\%, \tag{12}$$

where $D_{B4}$ and $D_{B0,4}$ are the simulated perturbed and baseline CBFi. A negative sensitivity indicates that the recovered perturbed CBFi decreases, whereas the simulated value increases, reflecting a poor CBFi recovery accuracy. A negative sensitivity indicates that the model responds inversely to changes in CBFi, while a sensitivity greater than 100% suggests that the model overreacts, amplifying the effect of CBFi changes.

*2.6 Optical and physiological parameters assumption errors*

Typically, when using DCS to extract CBFi, optical parameters ($\mu_a$, $\mu_s'$) and layer thicknesses are adopted from values reported in previous studies[4,42]. However, incorrect assumptions about these parameters can introduce significant inaccuracies in CBFi estimation. As the impact of parameter assumption errors on all three analytical models has already been examined in prior publications, we provide only a brief review of this topic in the present work and do not perform a quantitative analysis.

*2.7 Crosstalk from extracerebral BF*

The primary confounder in CBFi measurements using DCS is the blood flow in the extracerebral layer[81]. SBFi constitutes a significant portion of cerebral perfusion, while skull blood flow, though often overlooked, is also substantial due to its vascular networks and considerable thickness. We argue that assuming zero blood flow in the skull layer when using multi-layer DCS models to recover CBFi is not a good approximation. To investigate this, we varied the skull BFi (BBFi) while keeping blood flow in other layers constant, in order to study the correlation between recovered CBFi and BBFi variations across different analytical models.

In this section, we characterized the crosstalk raised by the variations in SBFi and BBFi separately. Specifically, we use the four-layer model to simulate $g_2$ data by varying the Brownian diffusion coefficient in the scalp layer varied across five levels: $2\times10^{-7}$, $6\times10^{-7}$, $1\times10^{-6}$, $1.4\times10^{-6}$, $1.8\times10^{-6}$ mm$^2$/s, while keeping the Brownian diffusion coefficients in other layers fixed at baseline values (Table 2). Similarly, to assess the effect of BBFi variations, the Brownian diffusion coefficient in the skull layer was set to five levels: 0, $4\times10^{-8}$, $8\times10^{-8}$, $1.2\times10^{-7}$, $1.6\times10^{-7}$ mm$^2$/s, while keeping baseline Brownian diffusion coefficients in the other layers. We use the fitting methods described in Sec. 2.4 to recover CBFi. To evaluate the sensitivity of the recovered CBFi to variations in extracerebral blood flow, we calculated sensitivities using Eq. (12), where $D_{B4}$ and $D_{B0,4}$ were replaced with the simulated SBFi and BBFi values under perturbed and baseline conditions, respectively.

*2.8 Time to results*

To evaluate the capability of different models for real-time CBF monitoring, we compared the computational time required by the semi-infinite, two-layer, and three-layer models using their respective fitting strategies. First, we compared ETLR fitting with single-exponential fitting for the semi-infinite model. Second, we assessed MD fitting versus SD fitting for the multi-layer models. Finally, we compared all fitting strategies across the three models and discussed their feasibility for real-time CBF monitoring.

**3 Results**

*3.1 Fitting methods*

In this section, we present the results of our comparison of different fitting methods for the three analytical models. The results are illustrated using box plots with error bars in Figs. 2, 3, and 4. In these plots, each box represents the interquartile range (25%–75%) of the recovered values (with a total of 20 samples for each parameter). The median is shown as a horizontal line inside the box, and the mean is indicated by a red point. The whiskers extend to 1.5 times the interquartile range (IQR), allowing identification of potential outliers.

Fig. 2 shows the results of semi-infinite ETLR fitting, comparing cases where $\beta$ is fixed versus $\beta$ treated as a fitting parameter. Both approaches significantly underestimate CBFi; however, fixing $\beta$ as a constant yields more stable estimates than allowing it to vary during fitting. The variability, indicated by larger error bars, increases with larger $\rho$ values. Additionally, $\beta$ can be accurately recovered using semi-infinite ETLR fitting.

Fig. 3 illustrates the performance of three fitting strategies for the two-layer analytical model: SD fitting with $\beta$ fixed, SD fitting with $\beta$ as a fitting parameter, and MD fitting. For SD fitting, treating $\beta$ as a fitting parameter results in large fluctuations particularly at larger $\rho$ values ($\geqslant$ 30 mm), leading to instability in the estimated BFi. In contrast, MD fitting yields relatively stable estimates of CBFi, SBFi, and $\beta$, although both CBFi and SBFi are slightly underestimated. Notably, SD fitting with $\beta$ fixed produces CBFi estimates that are closer to the ground truth than those obtained with MD fitting.

Fig. 4 presents results for the three-layer analytical model using both SD and MD fitting. The recovered CBFi and SBFi values are closer to the ground truth than those obtained from the semi-infinite or two-layer models, although there is a slight tendency to underestimate CBFi and overestimate SBFi. Among all methods, three-layer MD fitting provides the most accurate and stable estimates of CBFi and SBFi. In contrast, three-layer SD fitting with $\beta$ as a fitting parameter exhibits significant divergence and should be approached with caution due to concerns about the reliability of the recovered results.

Based on these results, we conclude that ETLR fitting with $\beta$ fixed is optimal for the semi-infinite model, SD fitting with $\beta$ fixed is preferable for the two-layer model, and MD fitting is best suited for the three-layer model. Therefore, we will adopt these respective fitting methods for further analyses in the following sections.

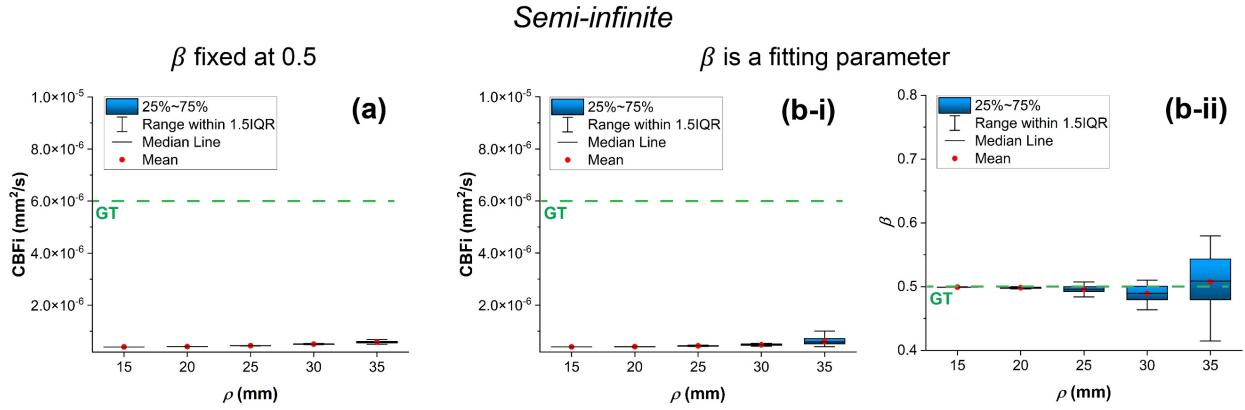

**Fig. 2** The semi-infinite model recovered CBFi at different $\rho$. (a) CBFi recovered using ETLR fitting, $\beta$ was fixed at 0.5. (b-i) CBFi recovered using ETLR fitting while $\beta$ was treated as a fitting parameter. (b-ii) recovered $\beta$ at different $\rho$. The green dashed lines with 'GT' represent parameter ground truths.

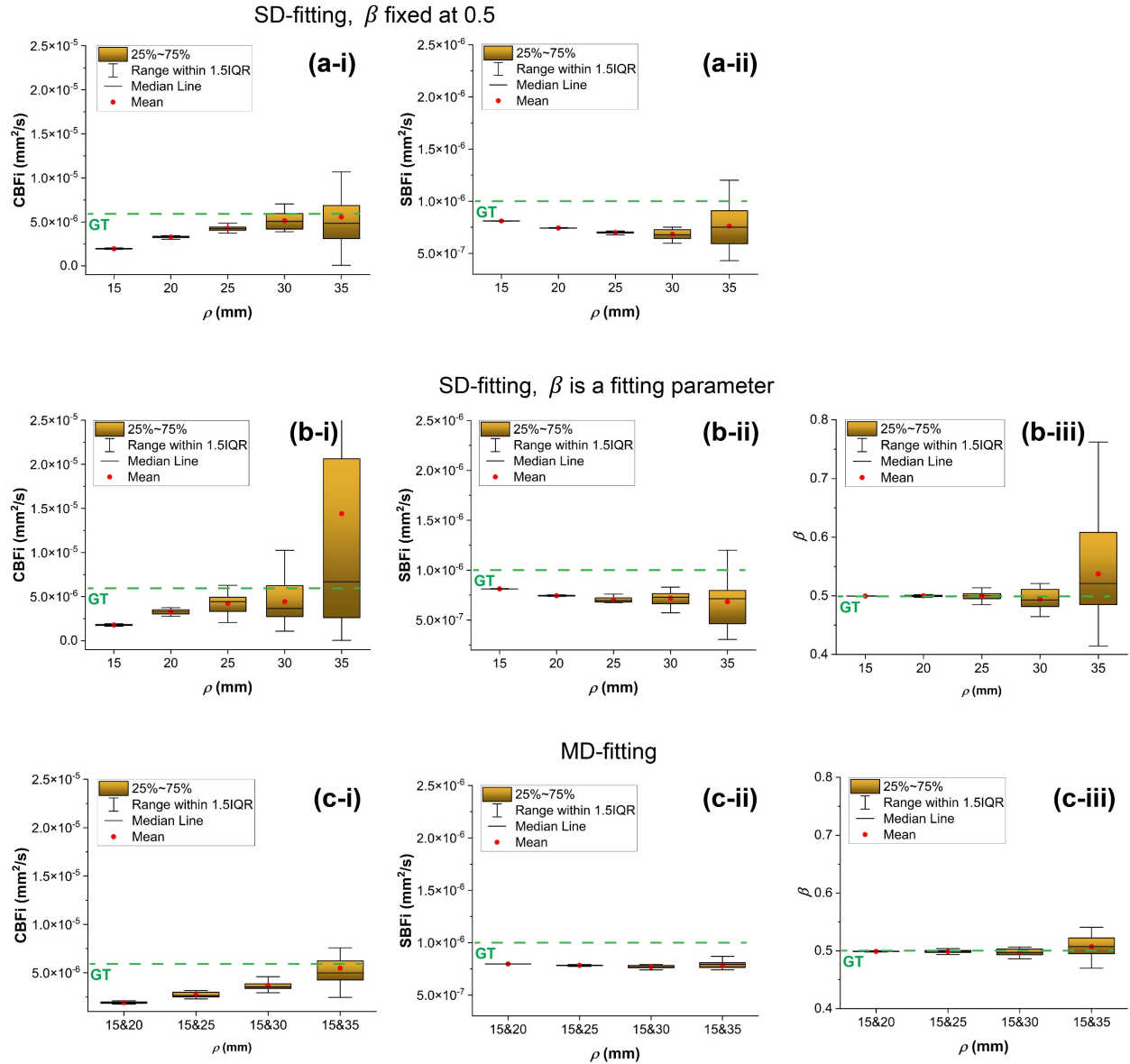

**Fig. 3** The two-layer model fitting results at different $\rho$. (a-i) and (a-ii) are the recovered CBFi and SBFi using SD fitting as $\beta = 0.5$. (b-i), (b-ii), and (b-iii) are the recovered CBFi, SBFi, and $\beta$ using SD fitting as $\beta$ was treated as a fitting parameter. (c-i), (c-ii), and (c-iii) are the recovered CBFi, SBFi, and $\beta$ using MD fitting. The green dashed lines with 'GT' represent parameter ground truths.

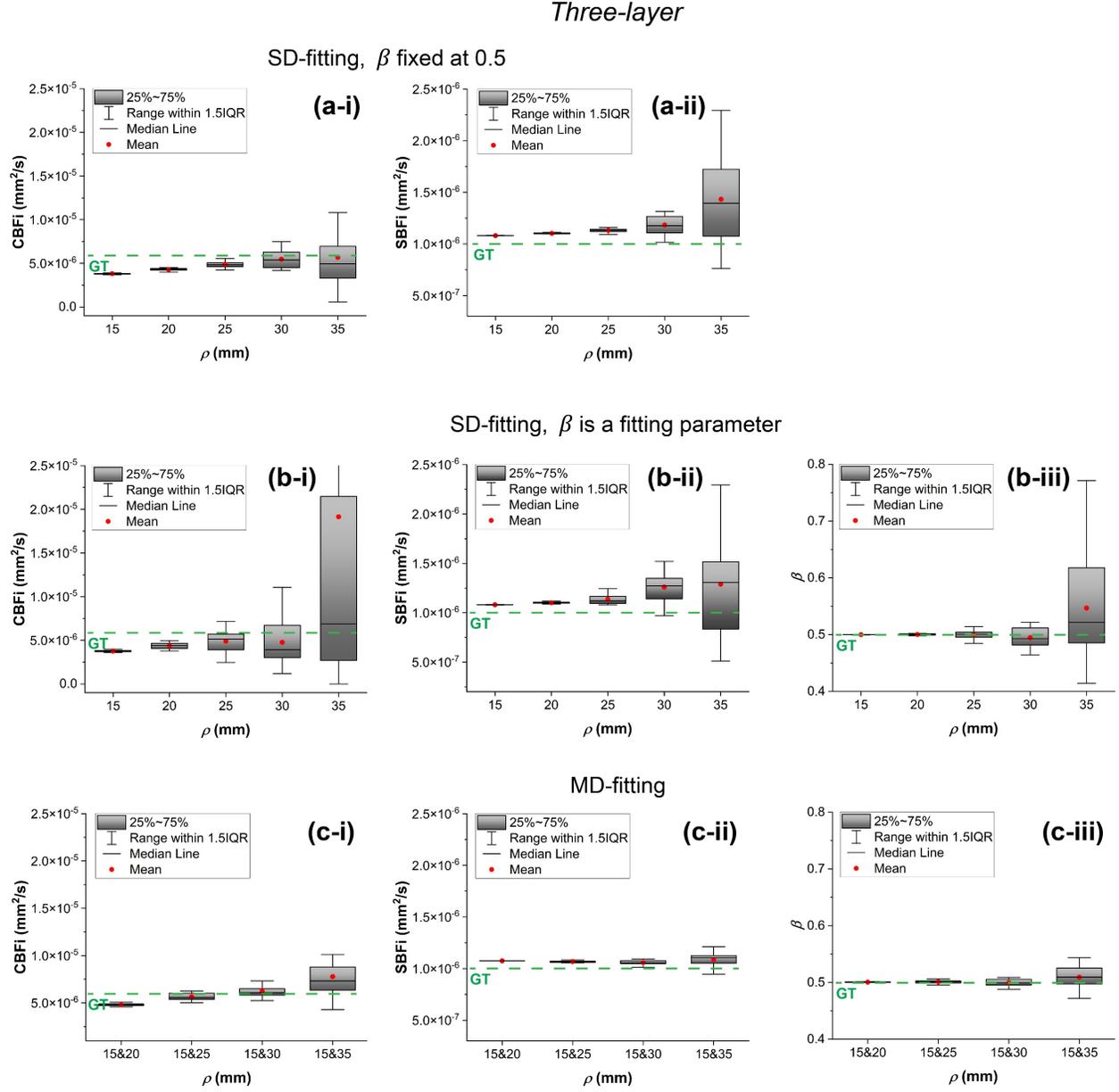

**Fig. 4** The three-layer model fitting results at different $\rho$. (a-i) and (a-ii) are the recovered CBFi and SBFi using SD fitting while $\beta = 0.5$. (b-i), (b-ii), and (b-iii) are the recovered CBFi, SBFi, and $\beta$ using SD fitting while $\beta$ was treated as a fitting parameter. (c-i), (c-ii), and (c-iii) are the recovered CBFi, SBFi, and $\beta$ using MD fitting. The green dashed lines with 'GT' represent parameter ground truths.

### 3.2 rCBFi recovery

In this section, we compare the recovered rCBFi ($\rho$ = 20, 25, 30, 35 mm) for the three analytical models using their corresponding fitting methods. Fig. 5(a) shows that rCBFi values recovered using semi-infinite ETLR fitting approach the ground truth as $\rho$ increases, although all estimates remain underestimated. In contrast, both two-layer SD fitting and three-layer MD fitting accurately recover rCBFi across all $\rho$ values, as shown in Fig. 5(b) and 5(c). However, at $\rho$ = 35 mm, the rCBFi estimates from two-layer SD fitting exhibit substantial fluctuations, making those results less reliable. Meanwhile, three-layer MD fitting demonstrates more consistent convergence at a larger $\rho$.

We also analyzed the performance of two-layer MD fitting for rCBFi recovery. Two-layer MD fitting tends to overestimate rCBFi across all $\rho$ values, with results presented in Fig. S3 of the Supplementary Material.

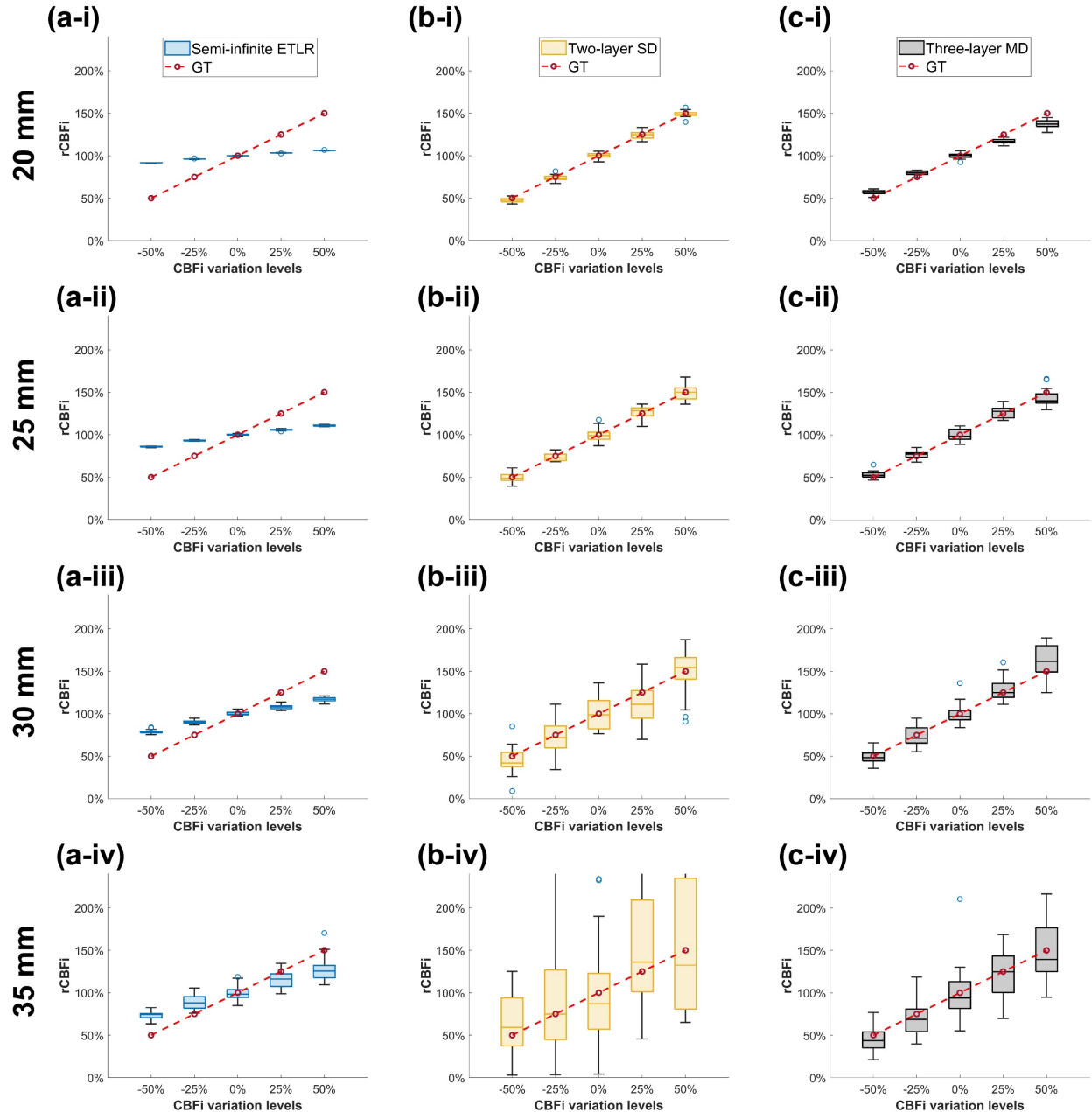

**Fig. 5** rCBFi recovered using semi-infinite ETLR fitting, two-layer SD fitting, and three-layer MD fitting at different $\rho$. The simulated CBFi perturbation has four levels, ±50% and ±25% compared with the baseline (6×10$^{-6}$ mm$^2$/s). We simulated 20 samples for each level at $\rho$ = 20, 25, 30, 35 mm.

*3.3 CBFi sensitivity*

To evaluate the sensitivity of different models to the changes in brain blood flow, we compared CBFi sensitivity using Eq. (12) on the same dataset used for rCBFi estimation. Fig. 6(a) shows that the CBFi sensitivity of semi-infinite ETLR fitting increases with larger $\rho$ values, reaching approximately 50% at 35 mm. Fig. 6(b) and 6(c) present the CBFi sensitivities for two-layer SD fitting and three-layer MD fitting, respectively. Both models exhibit comparable overall CBFi sensitivities, but differ in stability: at $\rho$ = 30 and 35 mm, the sensitivities show substantial fluctuations, indicating that variations in response magnitude could introduce errors in CBFi estimation. Notably, two-layer SD fitting occasionally exhibits negative sensitivity values, which, according to Eq. (12), imply a negative response to changes in CBFi.

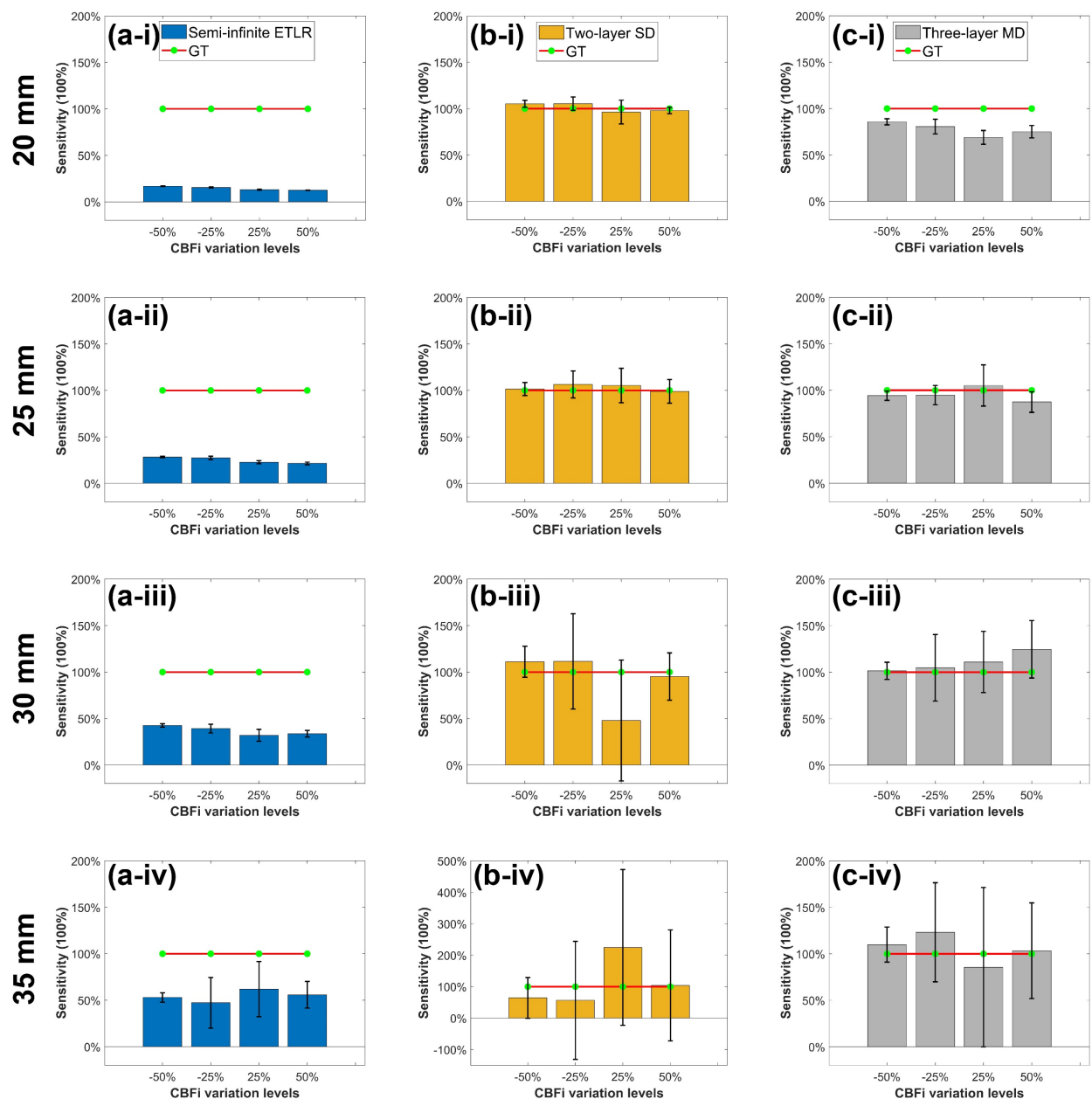

**Fig. 6** CBFi sensitivity using semi-infinite ETLR fitting, two-layer SD fitting, and three-layer MD fitting at different $\rho$ (20, 25, 30, 35 mm). The simulated CBFi perturbation has four levels, ±50% and ±25% compared with the baseline ($6\times10^{-6}$ mm$^2$/s). CBFi sensitivity was calculated using Eq. (12).

*3.4 CBFi sensitivity to extracerebral BFi changes*

In this section, we compare the CBFi sensitivity of the three analytical models to variations in extracerebral BFi. Five different levels of SBFi and BBFi were simulated using the four-layer head model, as introduced in Sec. 2.7. Figs. 7(a) and 7(b) show the simulated noise-free $g_2$ curves corresponding to variations in SBFi and BBFi, respectively. Changes in both SBFi and BBFi primarily affect the later portion of the $g_2$ curves. However, variations in SBFi result in more pronounced alterations in curve shape, as SBFi is substantially larger in magnitude than BBFi.

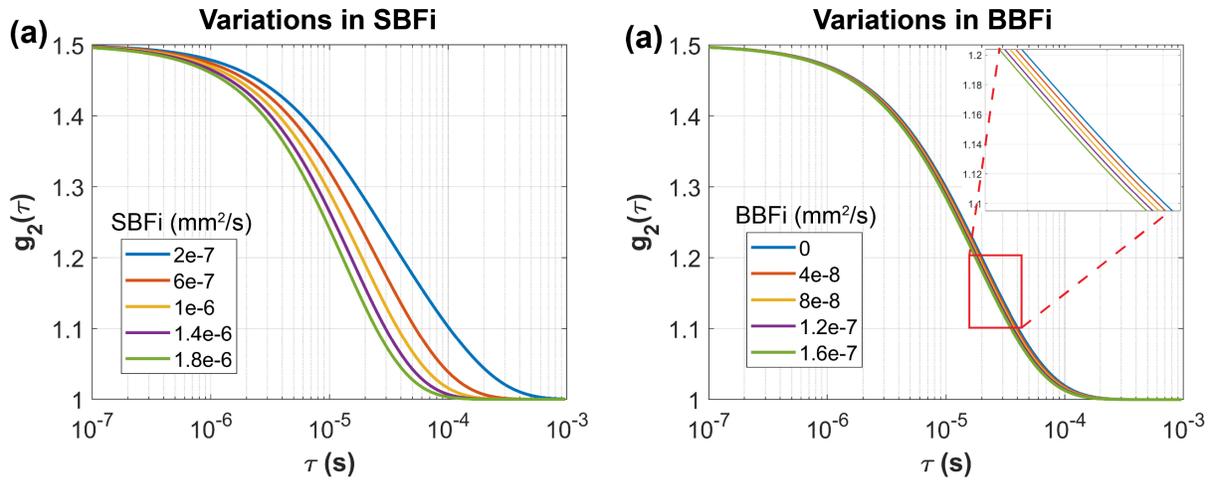

**Fig.** 7 The simulated $g_2$ curves with variations in SBFi and BBFi at $\rho$ = 30 mm.

Variations in SBFi

Fig. 8 illustrates the CBFi sensitivity of different models to variations in SBFi across various $\rho$ (20, 25, 30, and 35 mm). As shown in Fig. 8(a), the semi-infinite model exhibits positive CBFi sensitivity to SBFi variations at all $\rho$ values, with sensitivity decreasing as $\rho$ decreases. Consequently, the recovered CBFi increases with increasing $\rho$. For the two-layer model, sensitivities remain close to zero across all $\rho$ values, resulting in CBFi estimates that are relatively stable with respect to SBFi variations. However, the sensitivity fluctuates considerably at larger $\rho$, leading to instability in the estimated CBFi under SBFi variations. In contrast, the three-layer model shows predominantly positive sensitivities at $\rho$ = 20 and 25 mm, resulting in increased recovered CBFi values under SBFi variations. At $\rho$ = 35 mm, however, the sensitivity becomes negative, leading to decreased recovered CBFi. Among all tested distances, the three-layer model demonstrates relatively more stable CBFi recovery at $\rho$ = 30 mm.

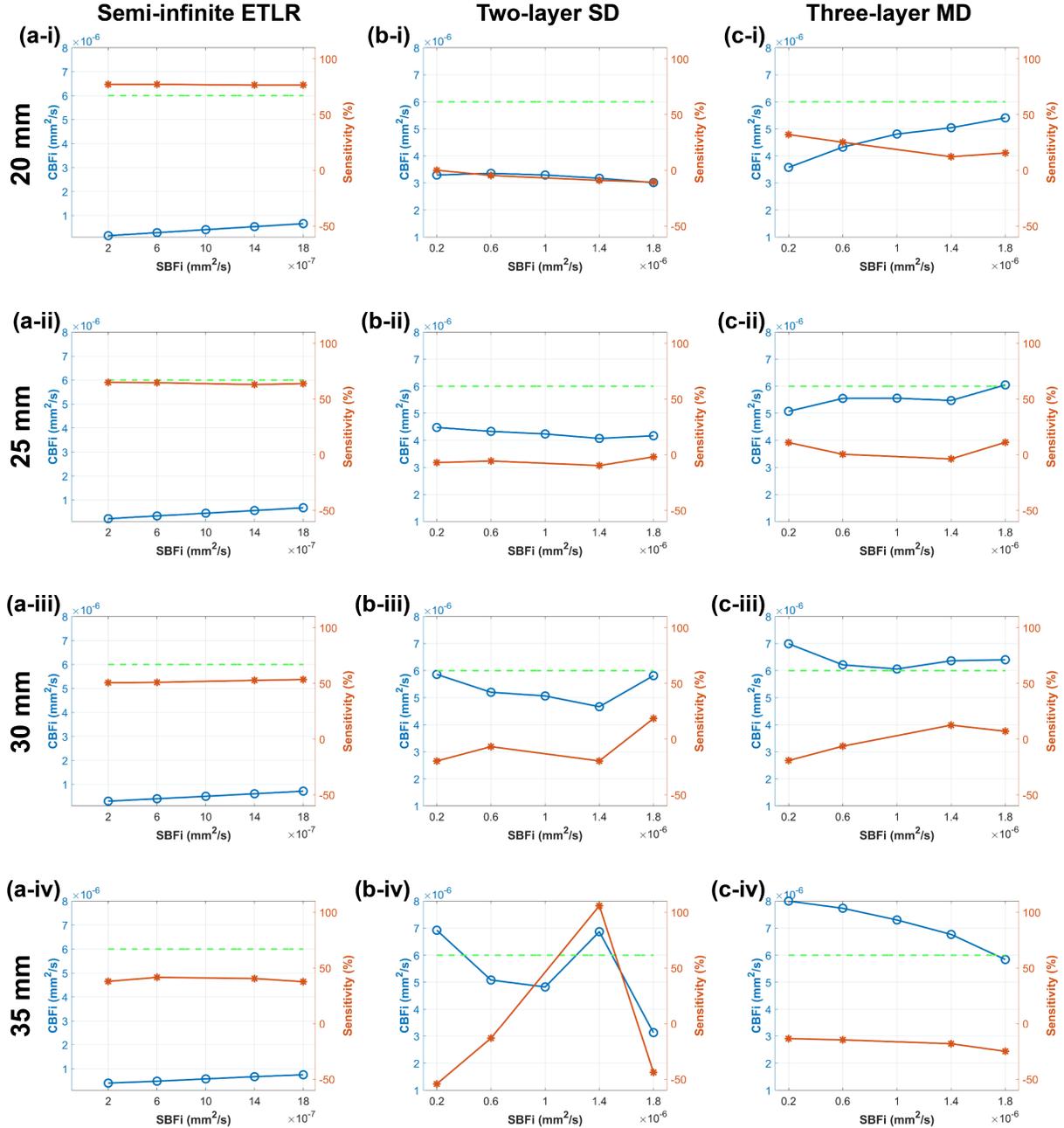

**Fig. 8** (a-i)-(a-iv): Semi-infinite ETLR fitting; (b-i)-(b-iv): two-layer SD fitting; and (c-i)-(c-iv): three-layer MD fitting recovered CBFi and CBFi sensitivity to SBFi variations across four different $\rho$ (20, 25, 30, 35 mm). The blue lines are the recovered CBFi at five different SBFi; the orange lines are the sensitivities calculated using Eq. (12); the dashed green line denotes the simulated CBFi ground truth. Note that for each variation, we simulated 20 samples, the plotted results are the median for the 20 samples to illustrate the tendency.

Variations in BBFi

Fig. 9 illustrates the CBFi sensitivity of different models to variations in BBFi across various $\rho$ (20, 25, 30, and 35 mm). As shown in Figs. 9(a) and 9(b), both the semi-infinite and two-layer models exhibit sensitivities close to zero, resulting in stable recovered CBFi across all $\rho$ values. In contrast, the three-layer model tends to display positive sensitivity at $\rho$ = 20, 25, and 30 mm, leading to an observed increase in recovered CBFi under BBFi variations.

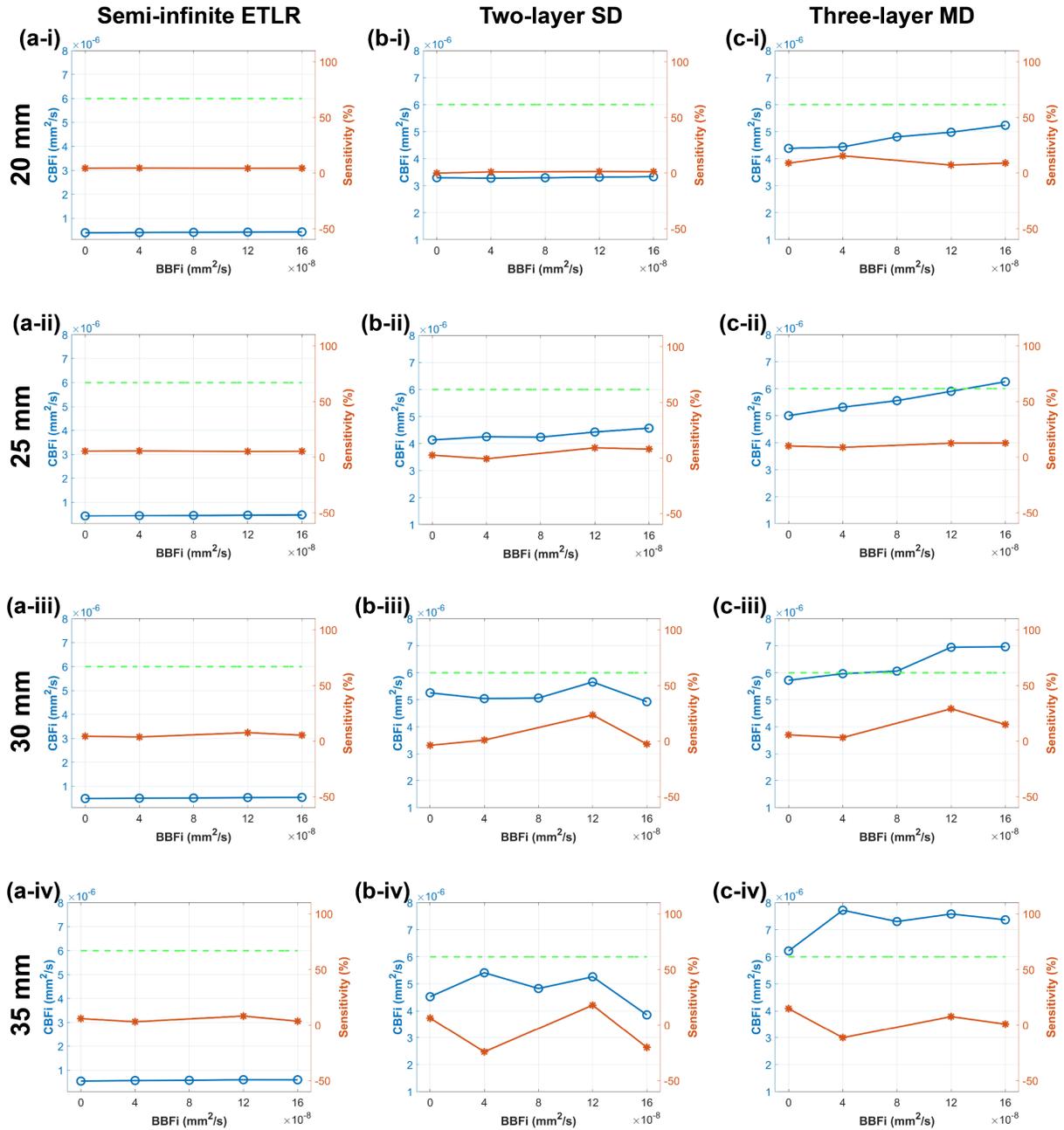

**Fig. 9** (a-i)-(a-iv): Semi-infinite ETLR fitting; (b-i)-(b-iv): two-layer SD fitting; and (c-i)-(c-iv): three-layer MD fitting recovered CBFi and CBFi sensitivity to BBFi variations across four different $\rho$ (20, 25, 30, 35 mm). The blue lines are the recovered CBFi at five different BBFi; the orange lines are the sensitivities calculated using Eq. (12); the dashed green line denotes the simulated CBFi ground truth. Note that for each variation, we simulated 20 samples, the plotted results are the median for the 20 samples to illustrate the tendency.

*3.5 Model parameters assumption error*

It is clear that errors in assumed model parameters can lead to inaccuracies in the estimated absolute CBFi. Several studies have quantitatively addressed this issue, listed in Table 1. Table 3 summarizes the parameters typically assumed for the three analytical models. For the semi-infinite model, the commonly assumed parameters are $\mu_a$ and $\mu_s'$. Previous reports[28,49] have quantitatively examined their influence, showing that $\mu_a$ has a positive effect on CBFi estimation, whereas $\mu_s'$ has a negative effect on CBFi estimation.

For the two-layer model, our previous work[82] demonstrated that brain $\mu_a$ has a positive relationship with CBFi estimation error, whereas $\mu_s'$ has a negative relationship. Moreover, the thickness of the extracerebral layer is positively associated with CBFi estimation error. Similarly, for the three-layer model, Zhao *et al.*[42] reported relationships consistent with those found for the two-layer model. It is worth noting that for both the two-layer and three-layer models, errors in assumed model parameters do not affect the recovered rCBFi, as reported by Pan *et al.*[82] and Zhao *et al.*[42], respectively.

*3.6 Comparison between semi-infinite ETLR and single-exponential fitting*

We compared and validated the effectiveness of using single-exponential fitting as a replacement for semi-infinite analytical fitting in rCBFi estimation. As shown in Fig. 10, single-exponential fitting demonstrates comparable accuracy and robustness in recovering rCBFi. Bland-Altman analysis reveals strong agreement between the two methods, suggesting that single-exponential fitting can serve as a simple and effective alternative for rCBFi estimation.

It is worth noting that we also applied the *lsqcurvefit* function, which is based on the Levenberg-Marquardt algorithm, to minimize the penalty function defined in Eq. (9) for single-exponential fitting. However, the rCBFi recovered using *lsqcurvefit* showed greater fluctuations and more underestimation compared to results obtained using the *fminsearchbnd* function (see Fig. S4 in the Supplementary Material). This finding suggests that *lsqcurvefit* may not be suitable for this task, and researchers should carefully consider the choice of optimization method when applying single-exponential fitting.

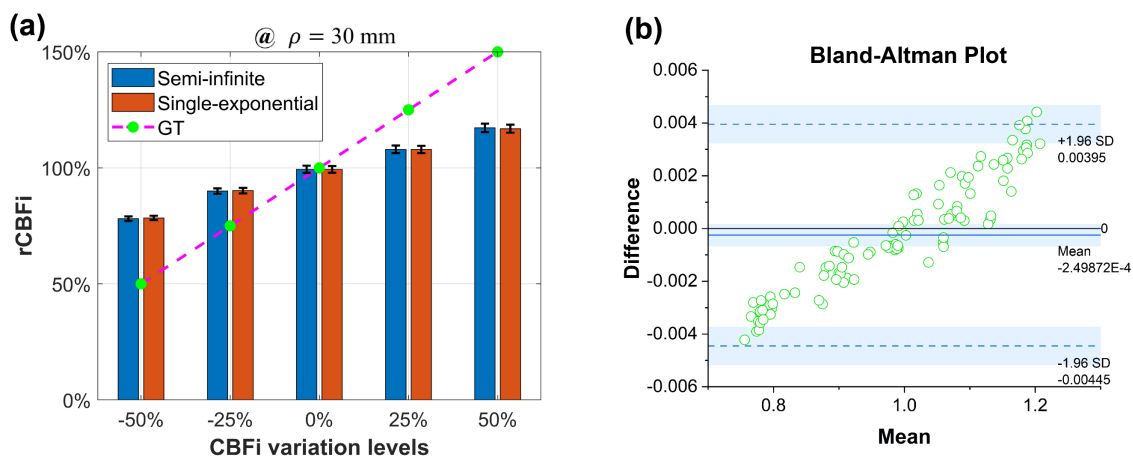

**Fig. 10** Comparison of rCBFi between semi-infinite analytical fitting and single-exponential fitting at $\rho = 30$ mm. *fminsearchbnd* function was used in MATLAB to minimize the penalty functions. (a) The bar represents the median of the recovered rCBFi, the error bar represents the 25[th] and 75[th] percentile range of the recovered rCBFi. (b) Bland-Altman plot of rCBFi recovered by the semi-infinite analytical fitting and single-exponential fitting at five different CBFi values, total 500 samples.

*3.7 Time-to-result comparison*

As defined in the penalty functions for the chosen fitting strategies for the semi-infinite (Eq. 7), two-layer, and three-layer (Eq. 10) models, the required predefined and fitted parameters are listed in Table 3. Table 4 summarizes the number of model parameters and the associated computational time required for fitting, based on the dataset used for rCBFi recovery, which includes a total of 500 data samples.

The semi-infinite and single-exponential models are clearly the fastest, processing 500 data samples in just 0.38 and 0.28 seconds, respectively. In contrast, the two-layer model requires 9,502.18 seconds, whereas the three-layer model takes 35,099.34 seconds using their respective fitting methods.

Table 3 Preknown and fitting parameters for each model using their respective fitting methods.

| Methods | Preknown parameters | Fitting parameters |
|---|---|---|
| Semi-infinite ETLR | $\beta, \mu_a, \mu_s'$ | $D_b$ |
| Single-exponential | $\beta$ | $\tau_c$ |
| Two-layer SD | $\beta, \mu_{a1}, \mu_{s1}', \mu_{a2}, \mu_{s2}', L_{extra}$ | $D_{b\_extra}, D_{b\_brain}$ |
| Three-layer MD | $\mu_{a1}, \mu_{s1}', \mu_{a2}, \mu_{s2}', , \mu_{a3}, \mu_{s3}', L_{scalp}, L_{skull}$ | $\beta, D_{b\_scalp}, D_{b\_brain}$ |

Table 4 Time cost for different models.

| Models | Number of Preknown parameters | Number of fitting parameters | Number of $g_2$ samples | Fitting methods | Optimization method | Time (s) |
|---|---|---|---|---|---|---|
| Semi-infinite | 3 | 1 | | ETLR | | 0.38 |
| Single-exponential | 1 | 1 | 500 | ETLR | *fminsearchbnd* | 0.28 |
| Two-layer | 6 | 2 | | SD | | 9502.18 |
| Three-layer | 8 | 3 | | MD | | 35099.34 |

**4 Discussion**

We systematically compared three DCS analytical models: the semi-infinite, two-layer, and three-layer models, in terms of CBFi sensitivity, accuracy of absolute CBFi and rCBFi recovery, ability to separate confounding signals from extracerebral layers, sensitivity to errors in assumed model parameters, and time-to-result. Through these comparisons, we identified several findings that, to our knowledge, have not been previously reported and are important considerations for researchers selecting or applying these models in different contexts.

First, we examined different fitting strategies for these models. For the semi-infinite model, we used ETLR fitting, as our primary goal was to assess performance in CBFi recovery. ETLR fitting can partially suppress contributions from superficial blood flow, thereby enhancing CBFi sensitivity. As shown in Fig. 2, the semi-infinite model consistently underestimates true CBFi, although the recovered CBFi increases with larger $\rho$ values. We also compared semi-infinite ETLR fitting with and without treating $\beta$ as a fitting parameter. Allowing $\beta$ to vary during fitting led to greater fluctuations in the results compared to fixing it as a constant. $\beta$ can be estimated by averaging the initial values of the $g_2$ curve at short $\tau$ and with long integration times. We recommend that researchers fix $\beta$ as a constant during semi-infinite ETLR fitting, as this approach yields more stable CBFi estimates, particularly since ETLR can reduce the SNR[58].

We also investigated different fitting methods for the two-layer and three-layer models. As shown in Fig. 3, two-layer SD-fitting with $\beta$ fixed provides more robust CBFi estimates across different $\rho$ values than fitting with $\beta$ as a free parameter. While two-layer MD-fitting offers the most robust estimation of all fitting parameters across the full range of $\rho$, it tends to deviate further from the ground truth due to model bias, particularly at smaller $\rho$. In contrast, the three-layer MD-fitting delivered the most robust and accurate estimates of all fitting parameters, with values closely matching the ground truth, as illustrated in Fig. 4. This demonstrates that the three-layer model can better capture the layered structure of the human head and is suitable for recovering both CBFi and SBFi simultaneously using MD-fitting.

Despite the promising performance of multi-layer models in CBFi recovery, it is important to recognize that they require more predefined model parameters for fitting, as summarized in Tables 3 and 4. When these parameters are inaccurately assumed, large errors in CBFi estimation can occur, particularly for the brain layer[28,42]. Therefore, researchers should carefully select the most suitable model based on their ability to reliably obtain the necessary parameters, as well as the fitting strategies discussed above.

However, in many DCS applications, rCBFi estimation is preferred[7,83], and the errors due to assumed model parameters are minimized, as reviewed in Sec. 3.5. Currently, the semi-infinite model is the most widely used approach for rCBFi estimation[9,10,12,15]. As shown in Fig. 5, the semi-infinite model's recovered rCBFi approaches the ground truth as $\rho$ increases and provides the most robust rCBFi estimates. These results indicate that the semi-infinite model can effectively recover rCBFi at larger $\rho$ values, while both the two-layer and three-layer models can effectively recover rCBFi across all $\rho$. Additionally, we validated single-exponential ETLR fitting and compared it to semi-infinite ETLR fitting at $\rho = 30$ mm, as shown in Fig. 10. Bland-Altman analysis indicated strong agreement between the two methods in recovering CBFi, and both demonstrated comparable accuracy and robustness in rCBFi recovery. These results suggest that it is feasible to use a simpler model for rCBFi estimation, without requiring prior knowledge of $\mu_a$ and $\mu_s'$. However, it is important to employ an appropriate optimization method, such as *fminsearchbnd*, and to avoid using *lsqcurvefit*, which showed inferior performance in this context.

As demonstrated in Fig. 6, CBFi sensitivity of the semi-infinite model increases with larger $\rho$, whereas the two-layer and three-layer models maintain CBFi sensitivity closer to the ground truth across all $\rho$. These results align with the findings shown in Figs. 2(a), 3(b), and 4(c): multi-layer models exhibit sufficient responsiveness to CBFi changes and can accurately recover absolute CBFi. However, as sensitivity fluctuates significantly at large $\rho$ due to decreased SNR, the multi-layer models' CBFi estimates also diverge considerably, potentially compromising reliability.

We further compared the three models' abilities to eliminate the influence of extracerebral layers (SBFi and BBFi). As shown in Figs. 8 and 9, the two-layer model most effectively accounts for BFi variations in extracerebral layers among all models. The semi-infinite model consistently shows positive sensitivity to SBFi and BBFi variations, indicating that its recovered CBFi is strongly correlated with blood flow changes in extracerebral layers. Zhao *et al.* reported that semi-infinite model-derived rCBFi is unaffected by extracerebral layer changes only when CBFi and SBFi change by the same fractional amount, which is a highly restrictive condition[42]. We believe that if a model exhibits non-zero sensitivity to changes in extracerebral blood flow, the recovered rCBFi is likely to be affected as well. Notably, the three-layer model appears more influenced by extracerebral BFi changes than the two-layer model (see Figs. 8(c) and 9(c)). For example, although BBFi is relatively small, the three-layer model shows positive sensitivity to BBFi changes, resulting in increased recovered CBFi as BBFi increases. This may be due to model bias when using a three-layer model to interpret a more complex anatomical structure, especially since MD-fitting can drive the optimal solution away from the ground truth, which researchers in this field should pay attention to. Another possible reason is the small discrepancies we observed between MC simulations and the analytical solution of the CDE, as also reported by Zhao *et al.*[55] This represents a drawback of MD-fitting that should be carefully considered.

Finally, we compared the time-to-result of different models using their corresponding fitting strategies to evaluate their suitability for CBFi monitoring. As listed in Table 4, the semi-infinite ETLR fitting and single-exponential ETLR fitting required significantly less time to process 500 data samples. In contrast, the two-layer and three-layer models were 25,000 and 92,367 times slower than the semi-infinite fitting, respectively. From a practical monitoring perspective, only the semi-infinite and single-exponential fittings are currently feasible for real-time applications. Encouragingly, deep learning (DL) methods have recently been introduced for DCS data processing and can greatly accelerate CBFi inference. However, most DCS-DL models have been developed based on the semi-infinite model, with only a few studies exploring multi-layer DCS models[82,84]. Based on our results, we believe multi-layer DCS models hold significant promise for accurate CBFi monitoring but will likely require advanced data processing techniques to be practically feasible.

Although we have comprehensively compared the three DCS analytical models, there are several limitations in our study that could be addressed in future work. First, for ETLR fitting, we constrained the lag time to $\tau \leq 30\ \mu s$ to simplify data analysis, whereas some previous studies have used criteria such as $g_2 \geq 1.2$ or $1.25$[22,75]. As shown in Fig. 1, ETLR selects different portions of the $g_2$ curve at different $\rho$ values, which may affect CBFi sensitivity; we did not perform a quantitative analysis of this influence. Second, as demonstrated in Sec. 3.1, we selected MD fitting for the three-layer model because it provides the most accurate estimates of both CBFi and SBFi. However, three-layer MD fitting tends to be sensitive to BFi changes in the extracerebral layers. To further evaluate the three-layer model's ability to separate extracerebral confounders, a comparison using three-layer SD fitting should also be conducted. Third, we generated our dataset using a four-layer slab model to represent the human head. Although this approach has been reported to introduce only small errors[6,57], future work should consider more complex and realistic head models to refine certain analyses. Additionally, in our simulations, the thicknesses of the extracerebral layers were kept constant. However, variations in layer thickness can influence brain sensitivity at different $\rho$ values. Therefore, future work should explore simulations incorporating varying extracerebral layer thicknesses to better understand these effects. Fourth, in adding noise to the simulated $g_2$ curves, we assumed a single photon count rate. Future studies should explore a broader range of noise levels to assess the robustness of different models under varied experimental conditions.

## 5 Conclusion

In conclusion, we systematically compared three CW-DCS models: the semi-infinite, two-layer, and three-layer models. We evaluated different fitting strategies, intrinsic CBFi sensitivity, accuracy of CBFi and rCBFi recovery, performance in separating extracerebral blood flow influence, sensitivity to errors in assumed parameters, and the feasibility of real-time CBFi monitoring. Our findings indicate that the semi-infinite model is suitable for robust rCBFi estimation at large $\boldsymbol{\rho}$ values, the two-layer model performs well for both CBFi and rCBFi estimation across all $\boldsymbol{\rho}$ ($\geq 20$ mm in this study), and the three-layer model is the most suitable for simultaneously extracting CBFi, SBFi, and rCBFi across all $\boldsymbol{\rho}$. Among the three models, the two-layer model demonstrated the greatest robustness to BFi variations in extracerebral layers. Both the semi-infinite fitting and single-exponential fitting are suitable for real-time monitoring tasks. We believe this work offers a valuable reference for researchers aiming to accurately and appropriately apply DCS analytical models for CBFi quantification.

**Disclosures**

The authors declare no conflicts of interest.

**Code and Data Availability**

Data underlying the results presented in this paper are not publicly available at this time but may be obtained from the authors upon reasonable request.

**Funding**

This work has been funded by the Engineering and Physical Science Research Council (Grant No. EP/T00097X/1): the Quantum Technology Hub in Quantum Imaging (QuantiC) and the University of Strathclyde.

**Author Contributions**

Mingliang Pan conceptualized the study, derived the theoretical models, collected the data and performed the analysis. Quan Wang and Yuanzhe Zhang contributed to data analysis and reviewed the final manuscript. David Li was responsible for funding acquisition, project administration, provided the laboratory resources, and also reviewed and edited the manuscript.

**Acknowledgments**

Mingliang Pan and Yuanzhe Zhang would acknowledge the support from China Scholarship Council. We would like to acknowledge Dr. Qianqian Fang's input on MC simulations using MCX.

We also acknowledge Dr. Erin M. Buckley's valuable suggestions on adding noise across different source-detector separations.

# Supplementary Material:

*1. DCS analytical models' derivation*

*Semi-infinite model*

For a semi-infinite homogeneous medium, the correlation diffusion equation can be expressed as:

$$\left(\frac{D}{v}\nabla^2 - \mu_a - \frac{1}{3}\mu_s' k_0^2 \alpha \langle \Delta r^2(\tau) \rangle \right) G_1(\mathbf{r},\tau) = -S(\mathbf{r}), \quad (5)$$

The analytical solution to Eq. (1) under the assumption of extrapolated boundary conditions, is given by:

$$G_1(\rho,\tau) = \frac{3\mu_s'}{4\pi}\left[\frac{e^{(-Kr_1)}}{r_1} - \frac{e^{(-Kr_2)}}{r_2}\right], \quad (6)$$

where $D_r = v/(3\mu_s')$ is the photon diffusion coefficient, $v$ is the speed of light in the medium, $\tau$ is the delay time, $\rho$ is the source-detector separation, $K^2 = 3\mu_a\mu_s' + {\mu_s'}^2 k_0^2 \alpha \langle \Delta r^2(\tau) \rangle$, $k_0$ is the wavenumber of light in the medium, $\mu_a$ is the absorption coefficient, $\mu_s'$ is the reduced scattering coefficient, $r_1 = \sqrt{\rho^2 + z_0^2}$, $r_2 = \sqrt{\rho^2 + (z_0 + 2z_b)^2}$, $z_0 = 1/(\mu_a + \mu_s')$, $z_b = 2(1 + R_{eff})/(3\mu_s'(1 - R_{eff}))$ with $R_{eff} = -1.44n^{-2} + 0.71n^{-1} + 0.668 + 0.064n$ being the effective reflection coefficient, defined by the ratio of the refraction indices inside and outside the medium (e.g., $n = n_0/n_{air}$, $n_0$ is the medium refractive index, $n_{air}$ is the air refraction index).

*Two-layer model*

For the two-layer model, we treat the tissue geometry as two slabs. The upper layer accounts for superficial tissues, and the under one represents deeper tissue. For example, when using the two-layer model to extract CBFi, we simplify human head structure as one layer to represent extracerebral tissue (scalp, skull), and another semi-infinite part to represent cerebral tissue (cerebrospinal fluid, grey matter, white matter, etc.). Following the analytical derivation process developed by Gagnon *et al.* (Ref. 4 in the main text), we assume an isotropic source incident at depth $z_0 = 1/(\mu_{a,1} + \mu_{s,1}')$, and scatters in each layer present independent Brownian diffusion motion. Then the CDE will be

$$[D_1\nabla^2 - \mu_{a,1} - 2\mu_{s,1}' k_0^2 D_{B,1}\tau]G_1^1(x,y,z,\tau) = -\delta(x,y,z-z_0) \quad 0 \leq z \leq l,$$
$$[D_2\nabla^2 - \mu_{a,2} - 2\mu_{s,2}' k_0^2 D_{B,2}\tau]G_1^2(x,y,z,\tau) = 0 \quad l \leq z, \quad (3)$$

where $j = 1,2$ refers to the layer indices, $D_j$, $\mu_{a,j}$, $\mu_{s,j}'$, and $D_{B,j}$ are the diffusion coefficient, absorption coefficient, reduced scattering coefficient, and Brownian diffusion coefficient in Layer $j$, respectively, $l$ is the thickness of Layer 1. The Fourier domain solution to Eq. (3) at the surface of Layer 1 is

$$\tilde{G}_1^1(s,z,\tau) = \frac{\sinh[\alpha_1(z_b + z_0)]}{D_1\alpha_1} \times \frac{D_1\alpha_1\cosh[\alpha_1(l-z)] + D_2\alpha_2\sinh[\alpha_1(l-z)]}{D_1\alpha_1\cosh[\alpha_1(l+z_b)] + D_2\alpha_2\sinh[\alpha_1(l+z_b)]}$$
$$- \frac{\sinh[\alpha_1(z_0-z)]}{D_1\alpha_1}, \quad (4)$$

where $\alpha_j^2 = (D_j s^2 + \mu_{a,j} + 2v\mu'_{s,j}k_0^2 D_{B,j})/D_j$, $v$ is the light speed, $z_b = 2D_1(1 + R_{eff})/(1 - R_{eff})$. The Fourier inversion of Eq. (4) is

$$G_1^1(\rho, z, \tau) = \frac{1}{2\pi} \int_0^\infty \tilde{G}_1^1(s, z, \tau) s J_0(s\rho) ds, \quad (5)$$

where $J_0$ is the zeroth order Bessel function of the first kind.

*Three-layer model*

Similar to the two-layer model, we treat the tissue geometry as three slabs. For the human head, Layer 1 refers to scalp with a thickness $l_1$, Layer 2 refers to skull with a thickness $l_2$, and Layer 3 represents brain with a semi-infinite geometry type, i.e. $l_3 \to \infty$. Under the same assumptions, the CDE using three-layer model will be

$$[\nabla^2 - 3\mu_{a,j}\mu'_{s,j} + 6k_0^2 {\mu'_{s,j}}^2 D_{B,j}\tau] G_1(\mathbf{r}, \tau) = -s_0 \delta(\mathbf{r} - \mathbf{r}'), \quad (6)$$

We solve Eq. (3) in Fourier domain, and we can obtain the Fourier domain electric field temporal autocorrelation function as

$$\tilde{G}_1^0(\mathbf{q}, z = 0, \tau) = \frac{\text{Numerator}}{\text{Denominator}}, \quad (7)$$

$$\text{Numerator} = s_0 z_0 \{\beta_1 D_1 \cosh[\beta_1(l_1 - z_b)][\beta_2 D_2 \cosh(\beta_2 l_2) + \beta_3 D_3 \sinh(\beta_2 l_2)]\}$$
$$+ s_0 z_0 \{\beta_2 D_2 [\beta_3 D_3 \cosh(\beta_2 l_2) + \beta_2 D_2 \sinh(\beta_2 l_2)] \sinh[\beta_1(l_1 - z_b)]\}, \quad (8)$$

$$\text{Denominator} = \beta_2 D_2 \cosh(\beta_2 l_2)[\beta_1(D_1 + \beta_3 D_3 z_0)\cosh(\beta_1 l_1)$$
$$+ (\beta_3 D_3 + \beta_1^2 D_1 z_0)\sinh(\beta_1 l_1)]$$
$$+ [\beta_1(\beta_3 D_1 D_3 + \beta_2^2 D_2^2 z_0)\cosh(\beta_1 l_1)$$
$$+ (\beta_2^2 D_2^2 + \beta_1^2 \beta_3 D_1 D_3 z_0)\sinh(\beta_1 l_1)] \times \sinh(\beta_2 l_2), \quad (9)$$

where $\beta_j^2(\mathbf{q}, \tau) = 3\mu_{a,j}\mu'_{s,j} + 6k_0^2 {\mu'_{s,j}}^2 D_{B,j}\tau + \mathbf{q}^2$, $j = 1, 2, 3$ refers to the layer indices, $D_j = v/(3\mu'_{s,j})$, $\mu_{a,j}$, $\mu'_{s,j}$, $l_j$ and $D_{B,j}$ are the diffusion coefficient, absorption coefficient, reduced scattering coefficient, layer thickness, and Brownian diffusion coefficient in Layer $j$, respectively, $v$ is the light speed. $z_0 = 1/(\mu_{a,1} + \mu'_{s,1})$, $z_b = 2D_1(1 + R_{eff})/(1 - R_{eff})$

The measured field autocorrelation function at position $\mathbf{r} = \{\boldsymbol{\rho}, z = 0\}$ on the surface of the first layer will be (the inverse Fourier transform)

$$G_1^0(\mathbf{r}, \tau) = \frac{1}{(2\pi)^2} \int d^2\mathbf{q} \tilde{G}_1^0(\mathbf{q}, z = 0, \tau) e^{-i\mathbf{q}\cdot\boldsymbol{\rho}}$$
$$= \frac{1}{2\pi} \int d\mathbf{q} \, \tilde{G}_1^0(\mathbf{q}, z = 0, \tau) q J_0(\rho q), \quad (10)$$

where $J_0$ denotes the zeroth order Bessel function of the first kind.

2. SNR analysis of the synthesized experimental data

We validated our noise model using experimental results obtained from measurements on a milk phantom. The milk (1.8% fat, semi-skimmed, purchased from ALDI grocery) was diluted with water at a ratio of 1:3. We developed a traditional DCS system consisting of a multimode fiber (600 $\mu m$ core diameter, NA = 0.22, Thorlabs M160L01) delivering a 30 mW, 785 nm laser (DL785-100-S, CrystaLaser) to the sample. Scattered light was collected using a single-mode fiber (785HP, THORLABS) connected to a single-photon avalanche diode (SPAD; ID101-SMF20, Swiss Quantum). Photon arrival times were recorded and processed by a time-tagger module (SPC-QC-104, Becker & Hickl GmbH) to compute the intensity autocorrelation function, $g_2$. We then applied the semi-infinite analytical model to fit the experimental data. Fig. S1 shows both the raw $g_2$ curves and the corresponding fitted (analytical) curves.

To validate the noise model, we conducted experiments at two $\rho$ (20, 30 mm). The observed photon count rates at 20 mm and 30 mm were 121.1 kcps and 29.5 kcps, respectively. We set the integration time to 1 second. For each $\rho$, we collected 100 $g_2$ curves to calculate the standard deviation at each time lag $\tau$. The observed step-like pattern in the results is due to the multi-tau configuration of the hardware correlator, where the bin width increases stepwise (6.145 ns for the first 16-channel and is tripled every 16-channel thereafter). As stated in the main text, as a validation, the parameters used in our noise model were all derived from experimental settings and observations.

Fig. S2(a) illustrates the noise level (standard deviation, $\sigma$, at each $\tau$) at $\rho$ = 20 mm and 30 mm, validating our theoretical noise model against experimental measurements. Fig. S2(b) presents the SNR computed from both the theoretical noise model and the experimental analysis, using Eq. (6) from the main text. Both the noise and SNR analyses demonstrate excellent agreement between the theoretical noise model and experimental results as reported in previous studies, confirming the reliability of the noise-adding method employed in this study.

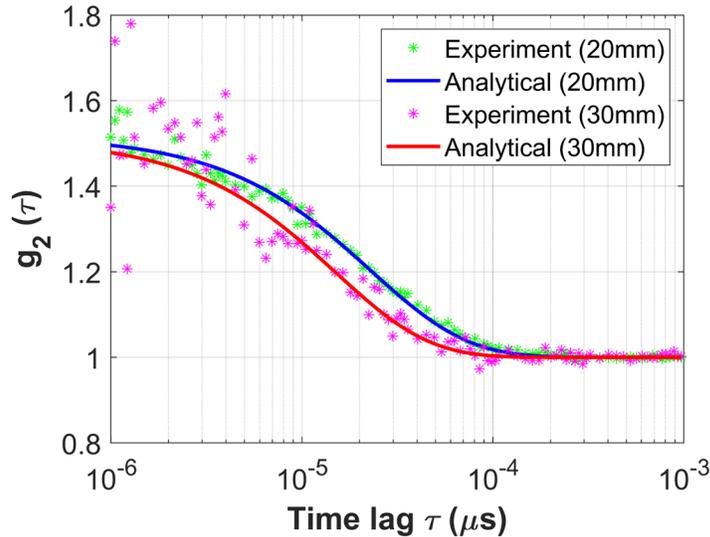

**Fig. S1** Noisy $g_2$ curves (scatter plots) alongside the fitted $g_2$ curves from the semi-infinite analytical model (solid lines) at $\rho$ = 20 mm and 30 mm.

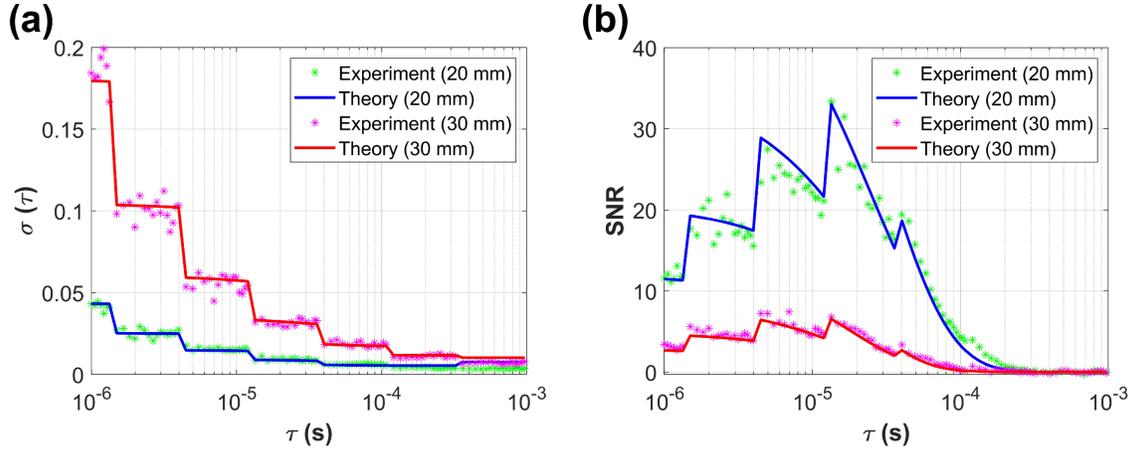

**Fig. S2** Validation of the noise model using the milk phantom experiment. (a) Standard deviation ($\sigma$) as a function of $\tau$ from the noise model, compared against experimental results at $\rho$ = 20 mm and 30 mm. (b) SNR calculated from the noise model, validated against experimental results. For the experimental analysis, each measurement was repeated 100 times to perform statistical analysis.

## 3. Two-layer MD fitting for rCBFi recovery

We performed two-layer MD fitting to estimate rCBFi. Compared to two-layer SD fitting, the two-layer MD fitting achieves better convergence at larger $\rho$ values (35 mm). However, the two-layer MD fitting tends to overestimate rCBFi across all $\rho$, suggesting that its CBFi sensitivity is higher than that of the SD fitting method, which may lead to inaccuracies in estimating CBFi changes.

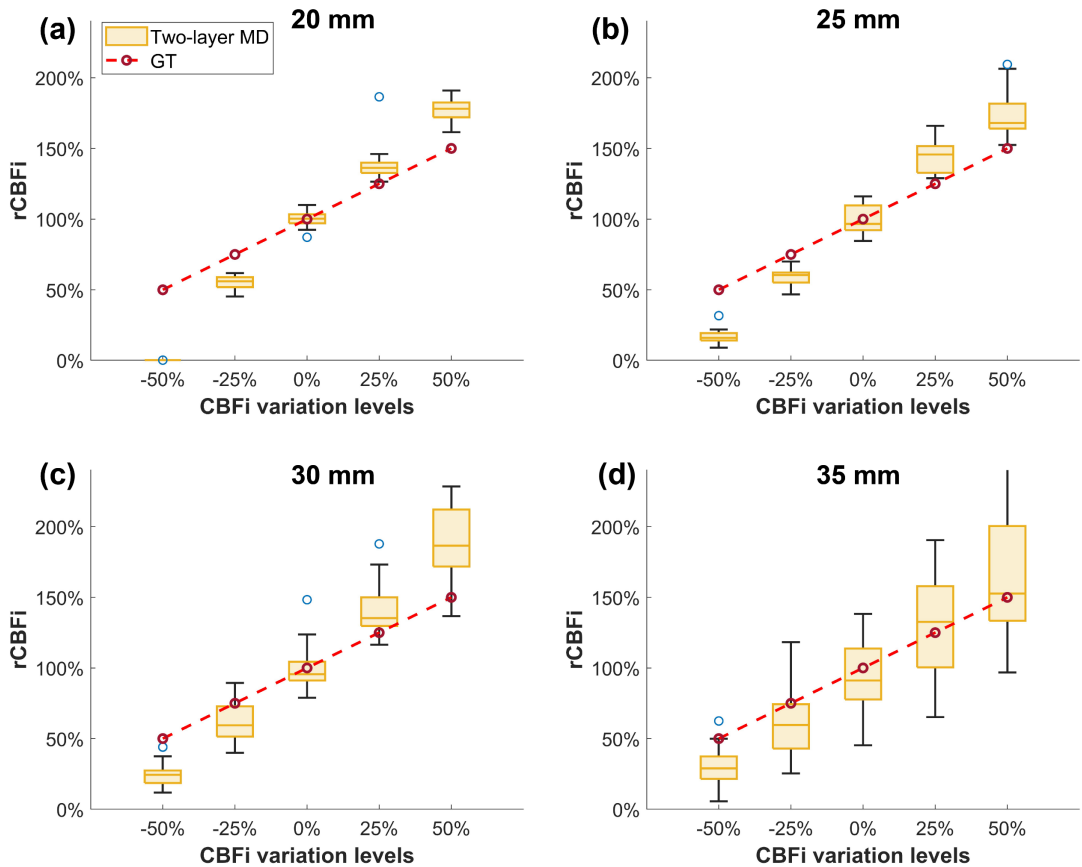

**Fig. S3** Two-layer MD fitting recovered rCBFi at different $\rho$ (20, 25, 30, 35 mm). The simulated CBFi perturbation has four levels, ±50% and ±25% compare to the baseline ($6\times10^{-6}$ mm$^2$/s).

## 4. Comparison between *fminsearchbnd* and *lsqcurvefit* for single-exponential fitting

We compared rCBFi recovered by single-exponential fitting using the *fminsearchbnd* and *lsqcurvefit* functions in MATLAB. As shown in Fig. S4, the rCBFi estimated with *lsqcurvefit* exhibits greater fluctuations and more pronounced underestimation compared to results obtained using *fminsearchbnd*.

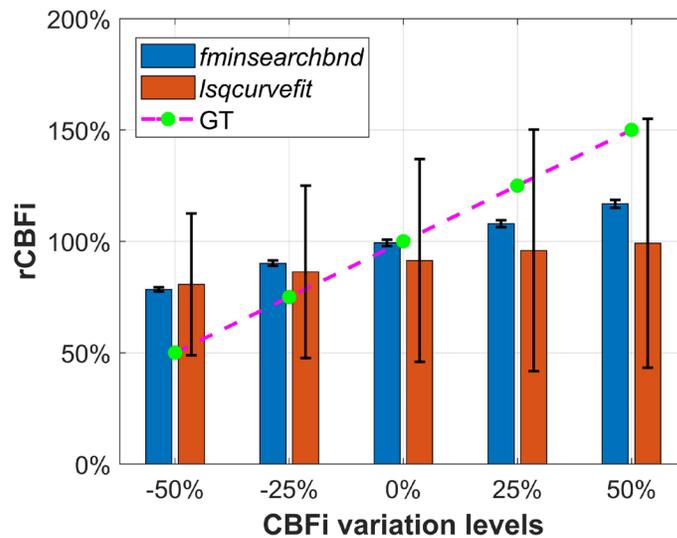

**Fig. S4** Single-exponential fitting recovered rCBFi using *fminsearchbnd* (blue) and *lsqcurvefit* (orange) functions. The bar represents the median of the recovered rCBFi, the error bar represents the 25th and 75th percentile range of the recovered rCBFi.